\documentclass[twocolumn,showpacs,preprintnumbers,amsmath,amssymb]{revtex4}

\usepackage[dvips]{graphicx}
\usepackage{dcolumn}
\usepackage{bm}

\begin{document}


\title{
CRITICAL PHENOMENA IN THE MAJORITY VOTER MODEL ON TWO DIMENSIONAL REGULAR LATTICES
}

\author{Ana L.\ Acu\~na-Lara}
\email{ana_lara@fisica.ugto.mx}
\affiliation{%
Departamento de Ingenier\'ia F\'isica,\ Divisi\'on de Ciencias e Ingenier\'ias,\\
Campus Le\'on de la Universidad de Guanajuato%
}

\author{Francisco Sastre}
\email{sastre@fisica.ugto.mx}
\affiliation{%
Departamento de Ingenier\'ia F\'isica,\ Divisi\'on de Ciencias e Ingenier\'ias,\\
Campus Le\'on de la Universidad de Guanajuato%
}

\author{Jos\'e Ra\'ul Vargas-Arriola}
\email{jr.vargasarriola@ugto.mx}
\affiliation{%
Departamento de F\'isica,\ Divisi\'on de Ciencias e Ingenier\'ias,\\
Campus Le\'on de la Universidad de Guanajuato%
}

\date{\today}

\begin{abstract}
In this work we studied the critical behavior of the critical point as function of the number of nearest 
neighbors on two dimensional regular lattices. We performed numerical simulations on triangular,
hexagonal  and  bilayer square lattices. Using standard finite size scaling theory we found that all cases fall
in the two dimensional Ising model universality class, but that the critical point value for the
bilayer lattice does not follow the regular tendency that the Ising model shows.

\end{abstract}


\maketitle

\section{Introduction}

The Majority voter (MV) model is a simple non-equilibrium Ising-like system that presents an order-disorder phase transition 
in regular and complex lattices. Numerical results for
the critical exponents, obtained on two-dimensional square
lattices~\cite{Oliveira92,Kwak2007,Wu2010} and
in three dimensional cubic lattice~\cite{AcunaLara2012}.  
are the same as those of the Ising model.
Those results seem to confirm the conjecture that non equilibrium models with up-down symmetry and spin flip dynamics
fall in the universality class of the equilibrium Ising model~\cite{Grinstein1985}. However, the critical
exponents obtained in numerical simulations on non-regular lattices~\cite{Campos2003,Lima2005,Pereira2005,Lima2006,Wu2010},
hypercubic lattices above four dimensions~\cite{Yang2008} and
two dimensional regular lattices with honeycomb and triangular geometries~\cite{Santos2011}
are different from the computed for the Ising model in the same geometries.

On the other hand the critical points reported for the MV model in Refs.~\cite{Oliveira92} and \cite{Santos2011}
seem to indicated a non monotonic behavior
as function of the number of nearest neighbors ($z$) in the lattice, the critical point for the square lattice ($z=4$) is bigger than those of the
triangular ($z=6$) and the honeycomb ($z=3$) lattices. This behavior clearly differs from the present in the Ising model for
the same geometries, even if we include the result for the bilayer square lattice ($z=5$)~\cite{Laosiritaworm2004}, where the inverse
critical temperature $\beta_c$ is a monotonic decreasing function on $z$.

We aim to clarify if the MV model belongs to the Ising model universality class in regular lattices and 
study the critical point dependence on the number of nearest neighbors in this model. 
In order to achieve our goals we evaluate the critical  points and the critical exponents for two dimensional
lattices in three different regular geometries:
the honeycomb, the triangular and the bilayer square lattice.

\section{Model and finite size scaling}
\label{model}

As mentioned above the MV model is an Ising-like system, in the sense that consists of a set of up-down "spins",
each one located on a lattice site that interact, in this work, with its nearest neighbors. The system evolves in
the following way: during an elementary time step, an spin $\sigma_i = \pm 1$ on the lattice is randomly picked up, and 
flipped with a probability given by
    \begin{equation}
    p(x)= \left\{ \begin{array}{ccc}
                \frac{1}{2}(1+x) & \mbox{for} & H_{i}\cdot\sigma_{i}< 0 \\
                \frac{1}{2}      & \mbox{for} & H_{i}=0    \\
                \frac{1}{2}(1-x) & \mbox{for} & H_{i}\cdot\sigma_{i}> 0
              \end{array}
    \right..%
    \label{transition}
    \end{equation}
Here $H_i$ is the local field produced by the nearest neighbors to the $i$~th spin and $x$ is the control parameter
(noise). A given spin $\sigma_i$ adopts the sign of $H_i$ (majority) with probability $(1+x)/2$ and the opposite sign of $H_i$ 
(minority) with probability $(1-x)/2$. In Refs.~\cite{Oliveira92,Wu2010,Campos2003,Lima2005,Pereira2005,Lima2006,Santos2011} 
an equivalent evolution rule is used, in those works a given spin $\sigma_i$ adopts the sign of the majority 
of its neighbors with probability $p$ and the sign of the minority with probability $(1-p)$. It is clear that with both evolution
rules the detailed balance condition is not satisfied.

The instantaneous order parameter $m_t$ is defined as
\begin{equation}
m_t =  \frac{1}{N} \sum_i \sigma_i,
\end{equation}
where $N$ is the total number of lattice sites.  
From here we can evaluate the moments of the order parameter 
as time averages 
    \begin{equation}
    \langle m^k\rangle =  \frac{1}{T-\tau} \sum_{t=\tau}^T |m_t|^k,
    \label{orderparameter}
    \end{equation}
where $\tau$ is the transient time and $T-\tau$ is the running time. 
The susceptibility is given by
     \begin{equation}
     \chi = N x \{\langle m^2\rangle-\langle m\rangle^2\}.
     \label{susceptibility}
     \end{equation}
We will use the method proposed in Ref.~\cite{Perez2005}, where two different cumulants are used for the
evaluation of the critical 
point, the fourth order-cumulant~\cite{Binder1981} (commonly known as 
Binder cumulant)
    \begin{equation}
    U^4=1-\frac{\langle m^4\rangle}{3\langle m^2\rangle^2},
    \label{cumulant4}
    \end{equation}
and the second order cumulant
    \begin{equation}
    U^2=1-\frac{2\langle m^2\rangle}{\pi\langle m\rangle^2}.
    \label{cumulant2}
    \end{equation}.

We assume that the same scaling forms used in the equilibrium models can be 
applied for the MV model. So, we will have a
free energy density given 
by the scaling ansatz
    \begin{equation}
    F(\epsilon,h,L)\approx  L^{-(d-\alpha)/\nu}f^0(\epsilon 
         L^{1/L},hL^{(\beta+\gamma)/\nu}),
    \label{scaling}
    \end{equation}
where $\epsilon=(x-x_c)$, $x_c$ is the critical point 
for the infinite system, $d$ is the dimension of the system, $f^0$ is a universal function,
$h$ is the symmetry-breaking (magnetic) field and $L$ is the linear dimension ($N=L^2$ for the
honeycomb and triangular lattices and $N=2L^2$ for the two-layer lattice). The 
parameters $\alpha$, $\beta$, $\gamma$ and $\nu$ are the 
critical exponents for the infinite system. 
From~(\ref{scaling}) the scaling forms for the thermodynamic 
observables can be obtained, with $h=0$ and one leading correction exponent, as
    \begin{eqnarray}
    m(\epsilon,L) &\approx&  L^{-\beta/\nu}(\hat{M}(\epsilon L^{1/\nu})+
          L^{-\omega} \hat{\hat{M}}(\epsilon L)),\\
    \chi(\epsilon,L) &\approx&  L^{\gamma/\nu}(\hat{\chi}(\epsilon 
          L^{1/\nu})+L^{-\omega} \hat{\hat{\chi}}(\epsilon L)),\\
    U^p(\epsilon,L) &\approx& \hat{U}^p(\epsilon L^{1/\nu})+
          L^{-\omega} \hat{\hat{U}}(\epsilon L).\label{scaling3}
    \end{eqnarray}
At the critical point $\epsilon=0$ we obtain the following set of equations that 
allow us to evaluate the critical exponents for small lattice sizes:
    \begin{eqnarray}
    m(L) &\propto& L^{-\beta/\nu}(1+ a L^{-\omega}), \label{beta}\\
    \chi(L) &\propto& L^{\gamma/\nu}(1+ b L^{-\omega}), \label{gamma}
    \end{eqnarray}
and
    \begin{equation}
    \frac{\partial U^p}{\partial x}\Bigl|_{x=x_c} 
        \propto L^{1/\nu}(1+c_p L^{-\omega}),
    \label{nu}
    \end{equation}
here $a,~b$ and $c_p$ are non universal constants. 
The critical point is evaluated taking in account that there are
differences between the crossing points 
in $U^2$ for different values of $L$, with respect to the corresponding
crossings evaluated for $U^4$.
The method take into account the 
correction-to-scaling effects on the crossing points.
We expand Eq.~(\ref{scaling3}) around $\epsilon=0$ to obtain
    \begin{equation}
    U^p\approx U^p_\infty+\bar{U}^p \epsilon L^{1/\nu} + 
        \bar{\bar{U}}^p L^{-\omega}+O(\epsilon^2,\epsilon L^{-\omega}),
    \label{crossing1}
    \end{equation}
where $p=2$ or $4$ and $U^p_\infty$ are universal quantities, but $\bar{U}^p$ and 
$\bar{\bar{U}}^p$ are non-universal. The value of $\epsilon$ where 
the cumulant curves $U^p$ for two different linear sizes $L_i$ and 
$L_j$ intercept is denoted as $\epsilon^p_{i,j}$. At this crossing 
point the following relation must be satisfied:
     \begin{equation}
     L_i^{1/\nu}\epsilon^p_{ij}+B^p L_i^{-\omega} =
     L_j^{1/\nu}\epsilon^p_{ij}+B^p L_j^{-\omega}.
     \end{equation}
Here $B^p=\bar{\bar{U}}^p/\bar{U}^p$. Combining for different cumulants ($q\ne p$) we get
    \begin{equation}
    \frac{x^{p}_{ij}+x^{q}_{ij}}{2}=x_{c}-(x_{ij}^{p}-x_{ij}^{q})A_{pq},
    \label{linealcrit}
    \end{equation}
where $A_{pq}=(B^p+B^q)/[2(B^p-B^q)]$ and $x^p_{i,j}=\epsilon^p_{i,j}+x_c $. 
Equation~(\ref{linealcrit}) is a linear equation that makes no reference to
$\nu$ or $\omega$, and requires as inputs only the numerically
measurable crossing couplings $x_{i,j}^p$. The intercept with the ordinate
gives the critical point location. Additional details of the method can be found
in Refs.~\cite{Perez2005} and \cite{AcunaLara2012}.

\section{Results}
\begin{figure}
     \begin{center}
     \includegraphics[width=7.50cm,clip]{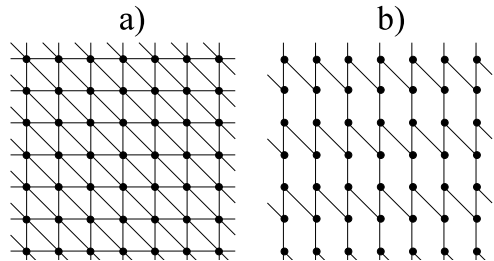}
     \caption{\label{tableaux}
     a) Triangular and b) honeycomb lattices
     use in our simulations. Both geometries were tested on the Ising model.
     }
     \end{center}
     \end{figure}

\begin{figure}
    \begin{center}
    \includegraphics[width=8.0cm,clip]{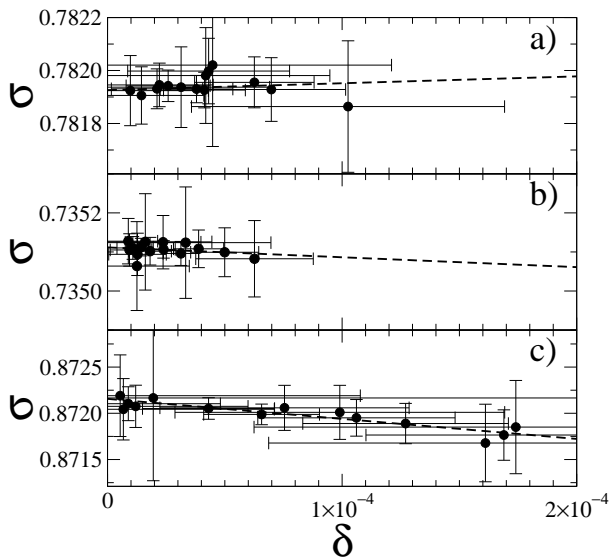}
    \caption{\label{criticalpoint} Evaluation of the critical 
    point for lattices with a) $z=6$, b) $z=5$ and c) $z=3$ nearest neighbors.
    The circles are the numerical data obtained with third order polynomial fits and the 
    dashed lines are the linear fits of Eq. (\ref{linealcrit}). The smaller $\delta$ values correspond
    to the larger system sizes.
    }
    \end{center}
    \end{figure}

We performed simulations on three different lattices with linear sizes
$L=24$, 28, 32, 36, 40 and 48. For the triangular and honeycomb lattices we
use the geometries shown in Figure~\ref{tableaux} with periodic boundary conditions.
For the bilayer square lattice we use a simple cubic lattice of size $N=2\times L\times L$ with periodic
boundary conditions along the $L$ direction and free boundary condition in the perpendicular direction,
in this way we get a system whose critical behavior is two-dimensional (at least for the Ising 
model~\cite{Laosiritaworm2004}) with $z=5$.
Starting with a random configuration 
of spins the system evolves following the dynamic rule explained in 
section~\ref{model}. 
In order to reach the stationary state we let the system evolves a transient time
that
varied from $2\times10^5$ Monte Carlo time steps (MCTS) for $L=24$ to $7\times10^5$ 
MCTS for $L=48$. Averages of the observables were taken over 
$2\times 10^6$ MCTS for $L=24$ and up to $7\times 10^6$ MCTS for $L=48$. 
Additionally, for each value of $x$ and $L$ we performed up to 200 independent 
runs in order to improve the statistics. 
Our simulations were performed in the $x$ ranges $[0.7785,0.7850]$, $[0.734,0.736]$ and $[0.869,0.875]$
for the triangular, bilayer and honeycomb lattices respectively.
For the evaluation of the critical points we use third order polynomial fitting 
for the cumulant curves. The estimation of the
critical points for the three cases are shown in Figure~\ref{criticalpoint}, where we use the 
notation $\delta=x_{ij}^4-x_{ij}^2$ and $\sigma=(x_{ij}^4+x_{ij}^2)/2$. 

The linear fits of Eq. (\ref{linealcrit}) give the following estimated for the critical points
$x_c=0.7271(1),~0.7351(1)$ and $0.8721(1)$ for $z=6$, 5 and 3 respectively.
The uncertainties in our results are one order of magnitude smaller with respect to the values reported
in Ref.~\cite{Santos2011}.
We also note that the smaller values in $\delta$ are around $1\time 10^{-5}$, as these values
correspond to the crossings between the largest sizes using in our simulation we can be sure that 
largest sizes will not improve significantly our results.

\begin{figure}
    \begin{center}
    \includegraphics[width=8.5cm,clip]{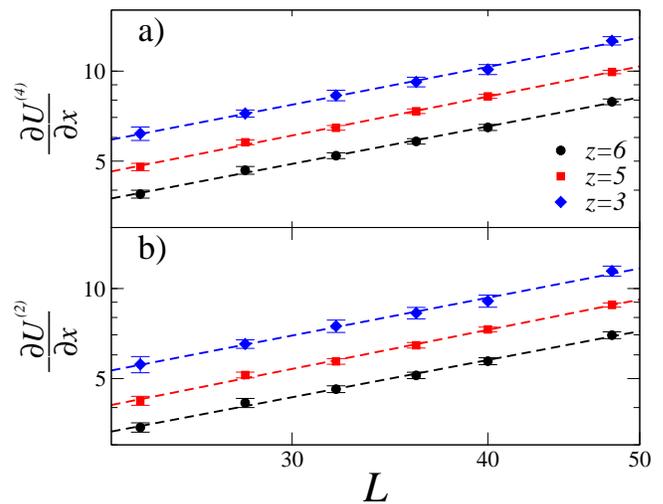}
    \caption{\label{nuexponent} (Color online)
    Log-log plot of the cumulant a) $U^4$ and b) $U^2$ derivatives at the critical point, 
    for $z=6$ (black circles), 5 (red squares) and 3 (blue diamonds).
    The dashed lines are power law fittings.
    }
    \end{center}
    \end{figure}

Once that we have the critical points we can evaluate the critical
exponents. For the evaluation of the critical exponent $\nu$ we use (\ref{nu}) 
with both cumulants $U^2$ and $U^4$. In Figure~\ref{nuexponent} 
we are showing the derivatives of the cumulants at the critical point for the three lattices.

The results for $1/\nu$ from the power law fits are given in Table~\ref{nuresults}. 
All results are in good agreement with the known value $1/\nu=1$ of the two dimensional Ising model.
    \begin{center}
    \begin{table}
    \caption{\label{nuresults} Estimates for $1/\nu$ obtained from the power law fitting of the cumulant derivatives
    at the critical point.}
    \begin{tabular}{@{}cccc@{}} \hline
    ~ & $z=6$ & $z=5$  & $z=3$ \\ \hline
    $U^2$ & 1.03(2) & 1.04(5) & 1.01(2) \\
    $U^4$ & 1.03(2) & 1.04(5) & 1.01(2) \\ \hline
    \end{tabular}
    \end{table}
    \end{center}

\begin{figure}
    \begin{center}
     \includegraphics[width=7.0cm,clip]{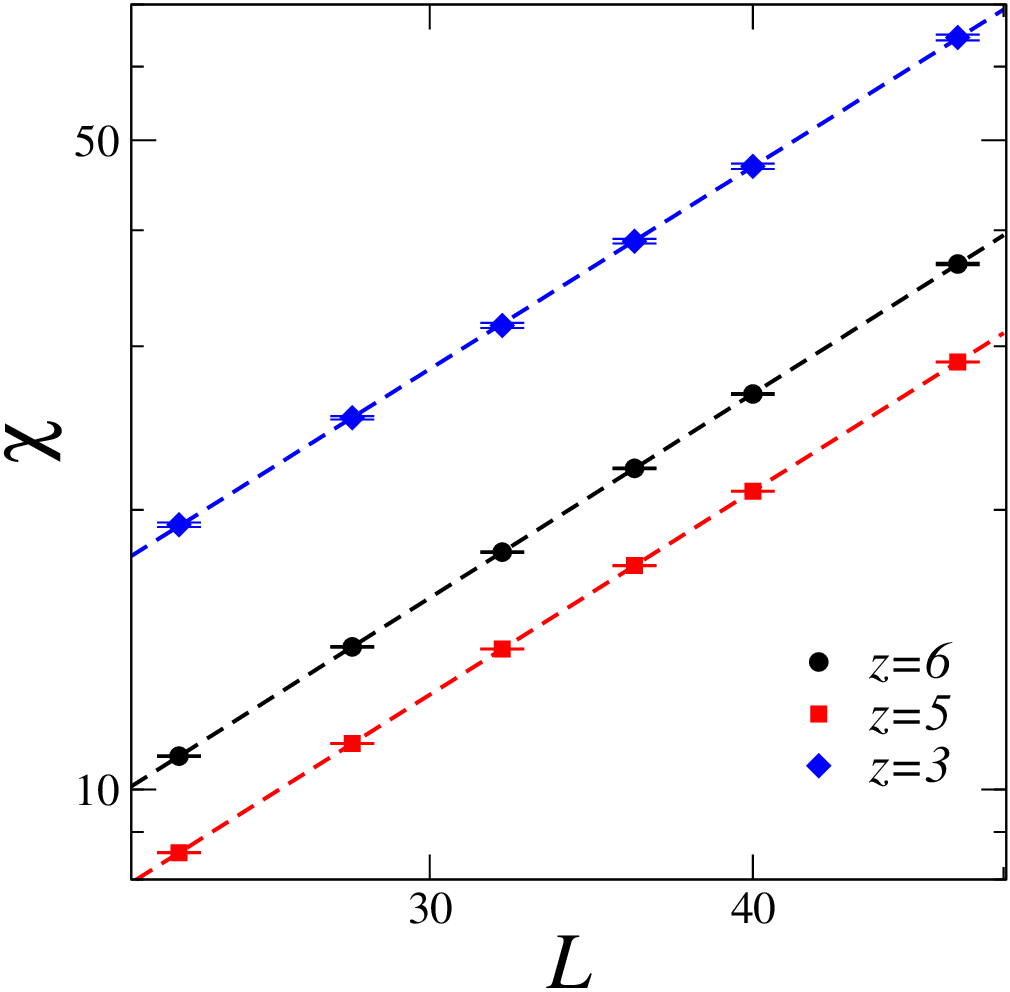}
     \caption{\label{gammaexponent} (Color online)
     Log-log plot of the susceptibility at the critical point
     for $z=6$ (black circles), 5 (red squares) and 3 (blue diamonds).
     The dashed line are power law fittings.}
    \end{center}
\end{figure}

For the critical exponent $\gamma$ we are using Eq.~(\ref{gamma}) to fit our data at the critical point (see Figure~\ref{gammaexponent}).
Our results are $\gamma/\nu=1.759(7),~1.756(9)$ and $1.755(8)$ for $z=6$, 5 and 3 respectively.  
Again the agreement is acceptable compared with the value $\gamma/\nu=7/4$ for the Ising model.

\begin{figure}
    \begin{center}
    \includegraphics[width=7.0cm,clip]{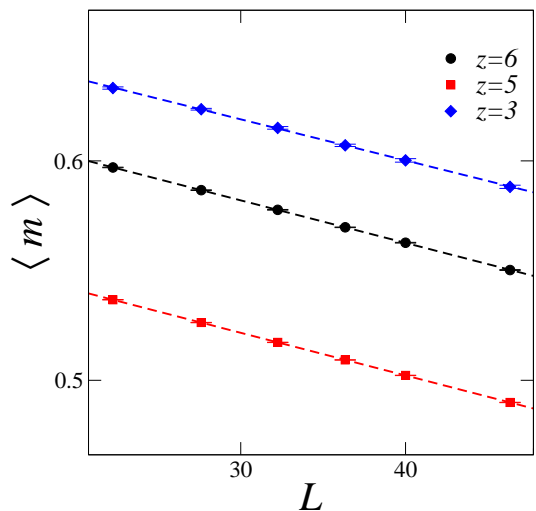}
    \caption{\label{betaexponent} (Color online)
    Log-log plot of the order parameter at the critical point for $z=6$
    (black circles), 5 (red squares) and 3 (blue diamonds).
    The dashed lines are power law fittings.}
    \end{center}
    \end{figure}

The fitting for the $\beta$ exponents are shown in Figure~\ref{betaexponent}. 
Our estimates are
$\beta/\nu =0.123(2),~0.123(3)$ and $0.123(2)$ for $z=6$, 5 and 3 respectively.
Those results also are in good agreement with $\beta/\nu=1/8$ for the Ising model.

It is important to point out that correction to scaling, via the exponent
$\omega$, were not necessary in the evaluation of the critical exponents
in the three geometries of this work. The same behavior was observed for the
MV model in square lattices~\cite{Oliveira92,Kwak2007,Wu2010}.
We summarized our results in Table~\ref{totales}, along with the reported values in Ref.
\cite{Santos2011} for the MV model in honeycomb and triangular lattices and the Ising model
in bilayer square lattice from Ref. \cite{Laosiritaworm2004}. We include also the results for
the  Rushbrooke-Josephson hyperscaling relation $d=(\gamma+2\beta)/\nu)$, which is satisfied by
our results. The fact that relation is compatible with $d=2$  by the bilayer square lattice
for the MV and Ising models indicate that both are two dimensional systems.

\begin{center}
\begin{table}[ht]
\caption{\label{totales} Critical values for the MV calculated in this work. For comparison we include the values
reported by Santos~{\it et al.}
for the MV model and the reported for the Ising model 
for the bilayer square lattice.}
\begin{tabular}{@{}ccccccc@{}} \hline
$x_c$       & $1/\nu$   & $\gamma/\nu$ & $\beta/\nu$ & $\frac{\gamma+2\beta}{\nu}$ & $z$ & model \\ \hline
0.7819(1)   & 1.03(2)   & 1.759(7)    & 0.123(2)    & 2.005(8)  & $6$ & MV \\
0.7351(1)   & 1.04(5)   & 1.756(9)    & 0.123(3)    & 2.002(11) & $5$ & MV \\
0.8721(1)   & 1.01(2)   & 1.755(8)    & 0.123(2)    & 2.001(9)  & $3$ & MV \\
0.772(10)   & 0.87(5)   & 1.59(5)     & 0.12(4)     & 1.83(5)   & $6$ & MV~\cite{Santos2011} \\
0.822(10)   & 0.87(5)   & 1.64(5)     & 0.15(5)     & 1.96(5)   & $3$ & MV~\cite{Santos2011} \\
---         & 1.00(1)   & 1.750(7)    & 0.126(7)    & 2.002(16) & $5$ & Ising~\cite{Laosiritaworm2004}    \\ \hline
\end{tabular}
\end{table}
\end{center}

The discrepancies between the results found in this work and the ones reported in Ref.~\cite{Santos2011}
can be explained because the critical point was not well enough evaluated, the uncertainities are ten time
biggers that the reported here. This would likely lead to a noticeable shift in the estimates of the 
critical  exponents.
Another source of discrepancies is the
incorrect definition for the susceptibility used in~\cite{Santos2011}, $\chi=N(\langle m^2\rangle-\langle m\rangle^2)$, instead
of (\ref{susceptibility}). The missing parameter $x$ it is necessary in order to have compatibility with the definition for
the susceptibility in the Ising model, $\frac{k_B T}{J}\chi= N(\langle m^2\rangle-\langle m\rangle^2)$, and this may be the cause
of the discrepancies in the $\gamma$ exponent. The same explanation can be applied to the works from
Refs.~\cite{Campos2003,Pereira2005,Lima2006,Wu2010}.
Finally we must mention that the numerical schemes applied in~\cite{Santos2011} can not be completely trusted, since
similar simulations for the Ising model on Archimedean lattices~\cite{Lima2011} by Lima {\em et al.}
were recently criticized by Malakis {\em et al.}~\cite{Malakis2012}.

\begin{figure}[ht]
    \begin{center}
    \includegraphics[width=8.0cm,clip]{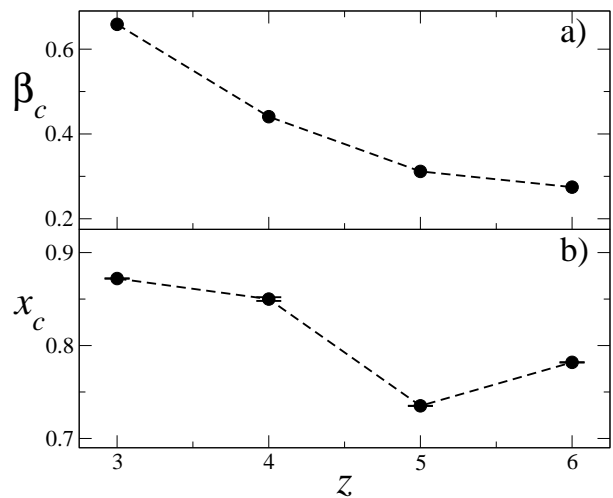}
    \caption{\label{puntocritico}
    Critical point behavior for the a) Ising and b) MV 
    models as function of the nearest neighbors number $z$. We observe that
    the MV model presents an anomaly at $z=5$.
    }
    \end{center}
    \end{figure}

Our results show also that the critical point is a monotonic decreasing 
function on $z$, with the exception from the bilayer case. In Fig.~\ref{puntocritico}
we can compare the difference in the behavior in the critical point of 
the MV and the Ising models~\cite{Laosiritaworm2004,Wannier1950}, the critical
point for the MV model in square lattice was taken from Ref.~\cite{Oliveira92}.

The unexpected behavior in the critical point for the bilayer square lattice does not affect the universality class,
but further studies are needed in order to check if there is a similar effect of this particular geometry
in other non equilibrium systems. 

\section{Conclusions}
The MV model on two-dimensional lattices with 3, 5 and 6 nearest neighbors belong
to Ising model universality class. Our simulations prove that 
the set of critical exponents for both models are consistent 
in two dimensional regular lattices.
The non equilibrium nature of the
MV model affects the critical point in the bilayer square lattice, but does not 
affect the universality class, at least for the static critical exponents.
It will be necessary to check if the critical dynamic exponents are the
same that those of the Ising model.

\section*{Acknowledgments}
A.\ L.\ Acu\~na-Lara and J.\ R.\ Vargas-Arriola thank Conacyt (M\'exico) for
fellowship support.
This work was supported by the DAIP (Universidad de Guanajuato, M\'exico)
through Grant 56-060.

\end{document}